\documentclass{aa}  

\usepackage{graphicx}
\usepackage{txfonts}
\DeclareRobustCommand{\ion}[2]{\textup{#1\,\textsc{\lowercase{#2}}}}
\newcommand{\livii}{\ensuremath{\mbox{\ion{$^7$Li}{I}}}}

\newcommand{\liratio}{\ensuremath{\mbox{$^6$Li\,} / \mbox{$^7$Li\,}}}
\newcommand{\LiI}{\ensuremath{\mbox{\ion{Li}{I}}}}

\newcommand{\KI}{\ensuremath{\mbox{\ion{K}{I}}}}
\newcommand{\NaI}{\ion{Na}{I}}

\newcommand{\ZnII}{\ion{Zn}{II}}

\newcommand{\HI}{\ion{H}{I}}
\newcommand{\hi}{\ion{H}{1}}
\newcommand{\htwo}{H$_2$}

\newcommand{\bvalue}{$b$-value}

\newcommand{\lya}{Ly$\alpha$}
\newcommand{\kms}{\ensuremath{{\rm km~s^{-1}}}}

\newcommand{\abund}[1]{\ensuremath{A({\rm #1})}}
\newcommand{\fuse}{{\em FUSE}}
\newcommand{\hst}{{\em HST}}

\newcommand{\lii}{Li\,{\sc i}}
\newcommand{\nai}{Na\,{\sc i}}
\newcommand{\liivii}{$^7$Li\,{\sc i}}
\newcommand{\livi}{$^6$Li\,}

\begin{document} 

   \title{Extragalactic  $^{85}$Rb/$^{87}$Rb  and $^6$Li/$^7$Li ratios  in  the Small Magellanic Cloud}


   \author{P. Molaro\inst{1,2}
          \and
          P. Bonifacio\inst{3,1}
          \and
          G. Cupani\inst{1}
          \and
          J. C. Howk\inst{4}
          }
          
\institute{INAF-OAT, Via G.B.Tiepolo 11, Trieste, I 34143, Italy\\
 \email{paolo.molaro@inaf.it}
         \and
        Institute  of Fundamental Physics of the Universe, IFPU, Via Beirut, 2, Trieste, I-34151, Italy\\
         \and
        GEPI, Observatoire de Paris, Université PSL, CNRS, 5 Place Jules Janssen, Meudon, 92190, France\\
     \and
         Department of Physics and Astronomy, University of Notre Dame, 225 Nieuwland Science Hall, Notre Dame, IN 46556, Indiana, USA\\
             }

   \date{Received September 15, 1996; accepted March 16, 1997}

 
\abstract
  {}
   { The line of sight toward Sk~143 (AzV 456),  an O9.5 Ib star in the Small Magellanic Cloud (SMC), shows significant
absorption from neutral atoms and molecules. We report a new study of this line of sight by means of  high-resolution  spectra obtained with the ESPRESSO spectrograph at the VLT of ESO. }
   {The absorption from neutral and ionized species is well characterized by a single component at v$_{hel}$ $\approx$ +132 \kms\ that was  modeled with the ASTROCOOK code.}
   {The  
  rubidium Rb~{\small I}  780.0 nm line is detected for the first time outside the Galaxy, and we derive [Rb/H]= -1.86 $\pm$ 0.09. As a result of the high resolution, the $^{85}$Rb and $^{87}$Rb  isotope lines are also exceptionally well resolved. The   $^{85}$Rb/$^{87}$Rb isotope ratio is  0.46, which
 is opposite of the meteoritic value of 2.43. 
 This  implies that Rb is made through  a  dominant   contribution of the   $r$-process, which is dominant for the $^{87}$Rb isotope. We also confirm the presence of \liivii\  670.7 nm   and set a limit on the isotopic ratio of $^6$Li/$^7Li$ < 0.1.}
   { The dominance of the  $^{87}$Rb isotope  implies that Rb is made through  a  dominant   contribution of the    $r$-process. At the  low metallicity  of  the cloud   of [Zn/H] = -1.28 $\pm$ 0.09 ,    neutron rich material   may have occurred   in rotating metal-poor massive stars. Moreover, the
  low     metallicity   of the cloud     leads to an absolute   Li  abundance  of  A($^7$Li) $\approx$ 2.2, which differs from  the expectation from  big bang nucleosynthesis. Because the gas-phase abundance is not affected by stellar depletion, the burning of Li inside the halo stars is probably not the solution for the cosmological $^7$Li problem.}


   \maketitle
 \titlerunning{Isotopes in the SMC}
%

\section{Introduction}

The formation and evolutionary history of galaxies can be reconstructed   by using the information contained in the elemental abundances observed in stars and in the interstellar medium. In this context, isotope ratios are particularly useful because they are  produced by different processes in different sources  that provide  physical information on the role of  asymtotic giant branch stars (AGB), novae, and supernovae
 \citep{kobayashi2011MNRAS.414.3231K}. However, the determination of isotopic ratios from quasar spectra requires data with very high quality, and isotopic determinations are only available for a small number of elements and are confined to the Milky Way.
 On the extragalactic scale, some information has been obtained for the $^{12}C/^{13}C$ ratio \citep{Levshakov2006, Muller2006,Carswell2011,welsh2020MNRAS.494.1411W}.
We conduct a first extragalactic study of the $^{85}$Rb/$^{87}$Rb and $^6$Li/$^7$Li   isotopic ratios along the line of sight toward Sk~143 by means of high-resolution  spectra acquired with the Echelle SPectrograph for Rocky Exoplanets and Stable Spectroscopic Observations, ESPRESSO,  installed at the incoherent combined Coudé facility of the Very Large Telescope of ESO \footnote{European Southern Observatory}.

  Sk~143 (AzV 456) is  an O9.5 Ib star in the Small Magellanic Cloud (SMC).
The line of sight toward the star  is unique in many respects. It shows  significant
absorption from neutral atoms and molecules at SMC velocities \citep{Cox2007,Welty2006,Cartledge2005}, together with a  rather strong reddening of E(B-V) = 0.36 or 0.33 \citep{welty2012ApJ...745..173W,jenkins2017ApJ...838...85J}. It also shows a peculiar Galaxy-type extinction curve that differs from those of other SMC stars. It shows also a rich set of molecules such as C$_2$, C$3
$, CN, and CH,  which reveal the presence of  a molecular cloud \citep{welty2013MNRAS.428.1107W}
Remarkably, \liivii\ 670.7 nm absorption was  detected,  along  with the possible presence  of \livi\   \citep{Howk2012Natur.489..121H}.  The only other extragalactic  detection of Li~{\small I} was  claimed in the spectra of  SN2014 in M82, but because this is a transient, it cannot be exploited further for the isotopic ratio \citep{Ritchey2015ApJ...799..197R}.

Rubidium (Rb, $Z = 37$) is a trace element  with a Solar System  abundance
of Rb/H = $2.8 \pm 0.2\times 10^{-10}$  as derived 
from meteorites \citep{lodders2021SSRv..217...44L}.
Earlier interstellar 
measurements in the Galaxy yielded upper limits \citep{Federman1985ApJ...290L..55F}, and the first      
interstellar detection was made based on 
observations of two heavily reddened Galactic stars in 
Cygnus \citep{Gredel2001A&A...375..553G}.  
In  stellar spectra,  the  resonance lines of
Rb~{\small I} at 780.0 and 794.7 nm are  blended
with atomic and/or molecular lines. 
Rubidium abundances 
were reported  for several  stars. They showed  that  [Rb/Fe] decreases with increasing metallicity, which is consistent   with the  prediction of chemical evolution models
 \citep{takeda2021AN....342..515T,caffau2021A&A...651A..20C}.

The production of Rb involves neutron
capture in both  $s$- and $r$-processes \citep{kappeler2011RvMP...83..157K}.  The
$s$-process occurs through  the main process in the He-shell
of low-mass AGB stars and  through the  weak process in the He- and C-burning layers of massive stars    \citep{shejeelammal2020JApA...41...37S}.  \citet{Garcia-hernandez2006Sci...314.1751G,Garcia-Hernandez2009ApJ...705L..31G} found  high Rb abundances   in  intermediate-mass AGB stars in the Galaxy  and  in the  Magellanic Clouds. This qualitatively agrees with theoretical models, which predict that  intermediate-mass AGB stars synthesize  substantial amounts of Rb \citep{vanraai2012A&A...540A..44V}.

 \livii\   and possibly also  \livi\ have been detected  toward Sk~143\citep{Howk2012Natur.489..121H}.  Sk~143 is relatively bright, and we made follow-up observations with the aim  to
 significantly reduce the statistical uncertainties in the  isotopic determination.  Measurements of 
\livii\ and \livi\ in
subsolar metallicity environments are quite   important to reconstruct their chemical evolutionary curve. \livii\ will help us to understand the cosmological Li problem, while  a significant abundance of \livi\  could come from $\alpha+\alpha$ fusion by Galactic or pre-Galactic cosmic rays or even from the effects of
non-Standard Model particles in the early Universe  \citep{pospelov2010ARNPS..60..539P}.

\begin{figure*}
 \includegraphics[width=18cm]{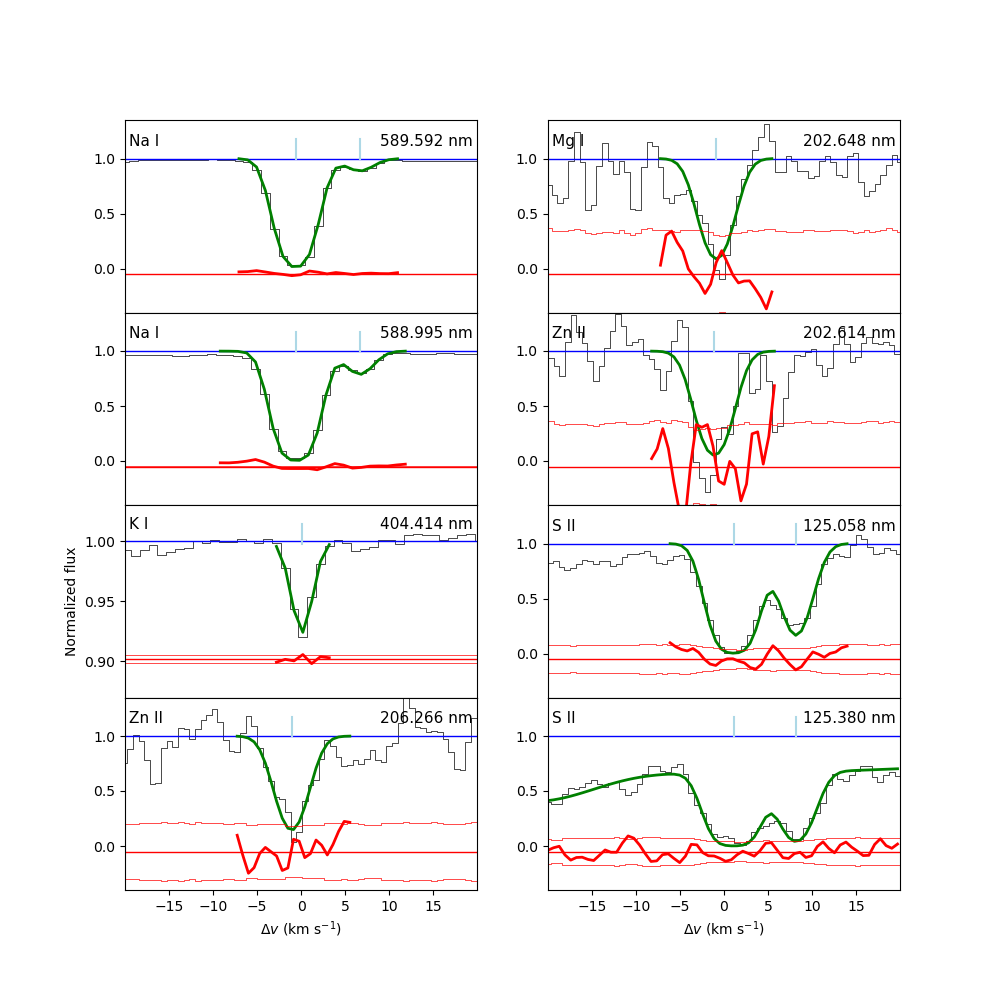}
    \caption{Absorption lines  for a representative set of elements  toward Sk~143 . The radial velocity correction reported in Table \ref{tab:full} was applied to the transitions. The data for the lines falling in the optical range were taken from our ESPRESSO observations, and those in the ultraviolet were taken from HST-STIS, the Space Telescope Imaging Spectrograph, acquired under program 9383 (PI: K. Gordon) and  reported in  \citet{Sofia2006}.
    The blue lines show the continuum, the green line shows the model that provides the best fit to the line absorption,  and the red line shows the  residuals.}
    \label{fig:sk143_all}
\end{figure*}
\begin{table}  
\vspace*{0mm}
 \caption{Journal of the observations.  } 
 \label{table1}
 \centering
 \scriptsize
\begin{tabular}{lccccc}
 \hline
   MJD$^{a}$ & t$_{exp}$&  ESPRESSO MODE&S/N\\
     $+50000$ &s && \\
 \hline
  \hline
59190.141431            & 3117 &HR21 & 56 &       \\
 59191.051152                 &3117  &HR21 &69 &       \\
 59191.091505                       & 3117 &HR21 &59 &       \\
 59191.130742                             & 3117 &HR21 &  58     & \\
 59191.170031                                   & 3117 &HR21 &  61    & \\
 59192.038096      & 3117 &HR21 &65 &       \\
 59192.146806            & 3117 &HR21 & 39&       \\
 59194.143836                  & 3117 &HR21 & 49&       \\
 59432.290725                        &3144  &HR42 & 43&       \\
 59432.328387                              & 3144 &HR42 & 37&       \\
 59433.189225 & 3144 &HR42 &77 &       \\
 59433.226818   & 3144 &HR42 &73 &       \\
59434.299063 & 3144 &HR42 & 48&       \\
59434.336597 & 3144 &HR42 &56 &       \\
59439.291095 & 3144 &HR42 & 83&       \\
 59458.221365& 3144 &HR42 &80&       \\
59458.261838 & 2803 &HR42 & 74&      \\
59459.276971 & 3144 &HR42 & 83&       \\
59459.317279 & 3144 &HR42 &71 &       \\
59461.160615 & 3144 &HR42 &80 &       \\
59461.198415 & 3144 &HR42 &84 &       \\
59462.109314 & 3144 &HR42 &80 &       \\
59462.148499 & 3144 &HR42 & 85&      \\
59462.202688 & 3144 &HR42 &95 &       \\
 59462.241639& 3144 &HR42 &98&       \\
 59462.280351& 3144 &HR42 &94&       \\
 59462.333779& 3144 &HR42 &87 &       \\
\hline
\end{tabular}
\tablefoot{ The time is the modified Julian date at the start of the observation. The signal-to-noise ratio is an average over all spectrum.
}
\end{table}

\section{Observations}\label{secA1}

Observations of  Sk~143 were   taken   with   ESPRESSO, the Echelle SPectrograph for Rocky Exoplanets and Stable Spectroscopic  Observations,  at the ESO Very Large Telescope. ESPRESSO has  two fibres, one fibre for the target and the other for the  sky,  with a diameter of $140\,\mu m$ that corresponds to a 1\farcs0 aperture in the sky \citep{Pepe2021A&A...645A..96P}. Twenty-seven observations   were performed in service mode with an individual exposure time of about $3150$\,s. 
The individual observations are reported  in Table \ref{table1}.
 The initial mode was 2x1 and subsequently  4x2 when this  became available.  In the latter mode, the binning of the CCD was  $4\times2$  pixels, that is, 4 pixels binned in the spatial direction and 2 pixels in the spectral direction, read in slow-readout mode. The  corresponding  resolution was R = $\lambda$ / $\delta \lambda$ $\approx$ 140.000 for both binnings.   

The ESPRESSO Data Reduction Software (DRS), version 3.0, was used for the data reduction. It includes bias and flat-fielding correction, wavelength calibration, and sky subtraction. The wavelength calibration combines a ThAr lamp  with a Fabry-P\'erot etalon. The sky subtraction used the sky spectrum observed by the second fiber of the instrument.  The observations were processed with the {\sc Astrocook} package \cite{2020SPIE11452E..1UC,cupani2023arXiv230510182C} to create a single one-dimensional spectrum. {\sc Astrocook} adopts a drizzling-like technique to combine several exposures without rebinning them individually; the different exposures were equalized to obviate possible discrepancies in the collected flux, and the errors from the DRS were propagated.

We also used archival STIS observations  acquired under program 9383 (PI: K. Gordon), which were reported previously \citep{Sofia2006}. The STIS observations were  taken with the E140H and
E230H gratings using the $0.20 \arcsec \times 0.09 \arcsec$ apertures with a 
resolution of  $R \approx 114,000$.

\section{The Sk~143 line of sight} \label{sec:sk143los}

The high-velocity gas  in absorption toward Sk~143 shows a heliocentric velocity of 132 as the systemic velocity of the SMC.
The hydrogen column density for the SMC component was derived from the   \lya\ absorption  using   Hubble Space Telescope (\hst)/STIS
E140H observations, which provided $\log
N(\hi) = 21.07 \pm 0.05$ \citep{Howk2012Natur.489..121H}.    The \htwo\ column was derived from \fuse\
data and provided a total hydrogen column along this sight
line of  $\log N({\rm H}) \equiv \log [N(\mbox{\HI})+2N({\rm
H_2})] = 21.46\pm0.04$ \citep{Cartledge2005}.  The gas has a high molecular fraction,
with $f({\rm H_2}) \equiv 2N({\rm H_2}) / N({\rm H}) = 0.6$.
The v$_{hel}$ = 132 \kms  absorption component
is  also visible in 21 cm emission, but the 21 cm profile contains
a significant amount of gas behind the star with far higher velocities.  
 The SMC is a disrupted galaxy with at least two major bodies: the SMC-Remnant, showing a characteristic velocity of $\approx$ 115 \kms; and the Mini-SMC with velocities of $\approx$  195 \kms \citep{mathewson1986ApJ...301..664M}. However, there are at least two other  substructures, one at  a radial velocity of $\approx  +134\pm 9$ \kms, as the the velocity of the gas  seen in front of Sk~143  \citep[see][and references therein]{molaro1990MmSAI..61..677M}. This  gas  belongs to a substructure in between the SMC Remnant, which is in the  southeast and not visible in the northwest at the position of  Sk~143, and the Mini-Small Magellanic Cloud, which lies behind the star. The   
 21 cm emission at much higher velocities  behind Sk~143  can be associated with the  Mini-SMC.

The main absorption features  toward Sk~143 at v$_{hel}$ $\approx$ + 132 \kms\  are displayed in Fig. \ref{fig:sk143_all}, together with the  model for the interstellar absorption and the residuals.
The absorption lines were modeled using {\sc Astrocook} \citep{2020SPIE11452E..1UC,cupani2023arXiv230510182C}.  The emission continuum was fit locally in the neighborhood of each absorption feature. A kappa-sigma clipping algorithm was used to iteratively reject the absorbed spectral bins. A Gaussian kernel was applied to the remaining bins to reconstruct the unabsorbed flux level. The results were visually inspected and manually corrected in a few cases by different observers to obtain the final reference continuum. Each Voigt profile was computed using the ionic mass and oscillator strength of the relevant species, and it was optimized for redshift $z$ and logarithmic column density $\log N$, while the Doppler-broadening parameter $b$ was kept constant. The instrumental profile was assumed to be Gaussian. For STIS,  we took the nominal STIS resolution from the STIS instrument handbook, namely the resolving power of two pixels, which is equivalent to R= 114000. For ESPRESSO, the resolving power of the SingleHR 4x2 mode was R= 130000 as measured 
 from the thorium and Fabry-Perot calibration frames \citep{Pepe2021A&A...645A..96P}.

The absorption from  neutral or single-ionised species is well characterized by a single component. The velocity dispersion is the same for heavy or light elements, and therefore, it is dominated by nonthermal motions. We verified that species with large differences in mass could be fit with the same choice of $b$, showing that the broadening was dominated by turbulence. We obtained the better fit with $b=1.6$ \kms{} for S {\sc ii}, Zn {\sc ii}, Mg {\sc ii}, and Na {\sc i}, while weaker lines were better fit with $b=1.0$ \kms (which confirms the choice of \citet{Howk2012Natur.489..121H}). 

A second component in the \nai\ lines that is redshifted by  about +7 \kms with respect to the main component was detected. This second component was also clearly visible in the high-resolution STIS spectra shown in \citet{Howk2012Natur.489..121H}. The column density of this redshifted component is much lower than that of the main component at 132 \kms and is not visible  in  the   other trace elements.

In Fig. \ref{fig:sk143_all} we show the main absorption features  toward Sk~143 in the SMC. The results are provided in  Table \ref{tab:full}. The values obtained from \citet{Howk2012Natur.489..121H}  are also reported in  Table \ref{tab:full}, and for the elements in common, they agree  remarkably well. A more exhaustive list  of elements  can be found in \citet{Howk2012Natur.489..121H}
and \citet{welty2016ApJ...821..118W}.

Zinc is especially interesting because it is a volatile element and   does not condense appreciably on interstellar dust grains, but  it closely tracks Fe in Galactic stars. For Zn, we adopted the  revised oscillator strength by \citet{kisielius2015ApJ...804...76K}. With the new values,  the  Zn metallicities are  lower by about 0.1 dex than the  determinations based on \citet{morton2003ApJS..149..205M}. Since the STIS E230H spectrum is rather noisy,  we kept the broadening value fixed at b = 1.6 \kms, as derived from all the strong lines,  
and   obtained  
 [Zn/H] = -1.28 $\pm$ 0.09, which agrees well with the value of [Zn/H] = -1.42 $\pm$ 0.33  derived by \citet{tchernyshyov2015ApJ...811...78T} by means of HST Cosmic Origin Spectrograph observations, and it is slightly below the [Zn/H]=$-$ 0.94 $\pm$ 0.09 by \citet{jenkins2017ApJ...838...85J}. This shows that the cloud has a low metallicity.
 \citet{mucciarelli2023A&A...671A.124M} studied hundreds of stars  in the SMC by grouping them according to their radial velocities and metallicities.
Interestingly, 37 giants form a   substructure called FLD-121, which is centered at RA 00:26 and DEC  –71:32. This is  not far from the  position of Sk~143 (RA 01:10 and DEC  –72:22).
The RV distribution of this stellar structure  peaks at RV $\approx +125$ \kms , and the  metallicity  distribution  ranges  from –0.8 dex to –2.2 dex;  20\% of the stars are more metal poor than –1.5 dex. FLD-121  is  quite different from the other 
structures, which show  metallicities about [Fe/H] $\approx$ -1 and likely
 formed in the first 1–2 Gyr of the life of the SMC \citet{mucciarelli2023A&A...671A.124M}. Giants are  a potentially different population and are on a different timescale than the gas would probe, but the
  oxygen abundance in the supergiant  Sk$~$143 was derived by \citet{evans2004ApJ...610.1021E}, who found [O/H] $<$ $-$ 0.76, which is also consistent with the metallicity of the interstellar cloud derived above, considering that some enhancement in the $\alpha$ elements is expected compared to the iron-peak elements.
 Thus, it is very likely that the gas seen toward Sk~143 at the  heliocentric velocity of $\approx$ 132 \kms  is associated with the  metal-poor stellar substructure FLD-121 of the SMC studied by \citet{mucciarelli2023A&A...671A.124M}.

\begin{figure}
        \includegraphics[width=\columnwidth]{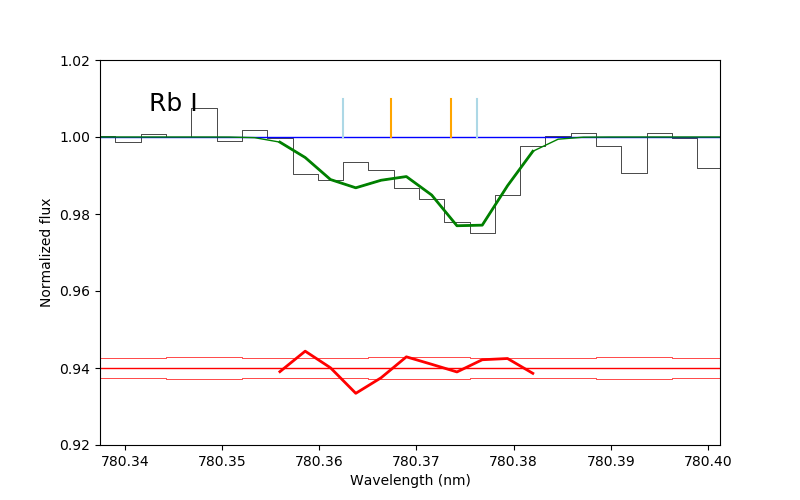}
    \caption{Spectrum of Sk143 around the  Rb~{\small I}  780.0 nm line. The positions of the hyperfine splitting of $^{85}$Rb (orange) and $^{87}$Rb (light blue) are indicated, and the green line shows the model. }
    \label{fig:fig_rubidium}
\end{figure}

\subsection{Rubidium and  $^{85}$Rb/$^{87}$Rb isotope ratio in the interstellar medium of the Small Magellanic Cloud}

The analysis of the Solar System abundances in meteorites provides estimates
of the fractional abundances of the isotopes $^{85}$Rb and $^{87}$Rb and of the main formation processes. The  Solar System $^{85}$Rb is composed of about 35\%   $s$- and
65\% of $r$-process. The weak $s$-process makes up about  30 \% 
of the $s$-process contribution \citep{Federman2004ApJ...603L.105F}.  For Solar System $^{87}$Rb, the fractions are inverted:   about 70\% is composed of   $s$- and about 30\% of $r$-process. For this isotope, only  
a minor contribution comes from the weak $s$-process. Rb is  a neutron densitometer
 because of its  role as a branch in the
$s$-process path    \citep{Lambert1995ApJ...450..302L}.     Low neutron densities favor $^{85}$Rb through the sequence  $^{84}$Kr$(n,\gamma)^{85}$Kr$(\beta) ^{85}$Rb, while high neutron densities favor $^{87}$Rb through the sequence 
$^{84}$Kr$(n,\gamma)^{85}$Kr$(n,\gamma)^{86}$Kr$(n,\gamma)^{87}$Kr$(\beta )^{87}$Rb.
 As a result of the
different relative contributions of the $s$- and $r$-processes
to the two isotopes, a measurement of the Rb isotope ratio is
a clue to the history of heavy-element nucleosynthesis.

 The  Rb~{\small I} 
spectrum for the interstellar gas toward Sk~143  is displayed in 
Fig. \ref{fig:fig_rubidium}.  The appearance of three components  arises from the 62 m\AA\ 
hyperfine splitting in $^{85}$Rb combined with the  
 hyperfine component in $^{87}$Rb.  
The Rb~{\small I} 
lines were fit to extract column densities for the isotopes $^{85}$Rb and 
$^{87}$Rb.  The relevant atomic data we used for input 
\citep{morton1991ApJS...77..119M,morton2000ApJS..130..403M} and the 
resulting  column densities ($N$) are given in Table \ref{tab:full}. 
The N({\rm Rb~{\small I}}) column density is  =  9.96 $\pm$ 0.08. With  $\log N(H) = 21.46 \pm 0.04$, its abundance  becomes [Rb/H] = -1.86 $\pm$ 0.09 without a correction for ionization. 
Insights into the absolute Rb abundance can be obtained through a comparison with potassium. The  K~{\small I}   abundance is obtained from the  K~{\small I} 404.414 nm  shown in Fig \ref{fig:sk143_all}. We derived a logarithmic N({\rm K~{\small I}})   column density of 12.57$\pm 0.1$, which  agrees well with the 12.61$\pm 0.0.2$  derived by \cite{Howk2012Natur.489..121H}, but the error is slightly larger, and with the 12.65 $\pm 0.1$ derived by \citet{welty2016ApJ...821..118W}.
Rb~{\small I} and K~{\small I} are trace interstellar species, and corrections for ionization are  required to estimate the  elemental 
abundances. The ionization 
potentials for Rb~{\small I} and K~{\small I} 
are similar, 4.18 and 4.34 eV, respectively,  and 
the depletion factors can be assumed to be similar 
for alkali  metals \citep{welty2001ApJS..133..345W,knauth2003ApJ...586..268K}. 
   The $A_g({\rm Rb})/A_g({\rm K})$ ratio is given by

\begin{equation}
\frac{A_g({\rm Rb})}{A_g({\rm K})} = 
\frac{N({\rm Rb~{\small I}})}{N({\rm K~{\small I}})}
\frac{G_{\rm Rb~{\small I}}}{G_{\rm K~{\small I}}}
\frac{\alpha_{\rm K~{\small I}}}{\alpha_{\rm Rb~{\small I}}},
\end{equation}

\noindent where $G$(X) are the photoionization rates  and $\alpha$(X) are the rate coefficients 
for radiative recombination.  
Following \cite{Federman2004ApJ...603L.105F}, the photoionization rates for 
Rb~{\small I} and K~{\small I} are $3.42 \times 10^{-12}$ and 
$8.67 \times 10^{-12}$ s$^{-1}$, respectively, and  $G_{\rm Rb~{\small I}}/G_{\rm K~{\small I}}$ = 0.3944. The 
  radiative recombination coefficients of Rb and K 
 are about the same, and therefore, the ratio $\alpha_{\rm K~{\small I}} / \alpha_{\rm Rb~{\small I}} $ is $ \approx$ 1.  
 Thus, from a measured  N$({\rm Rb~{\small I}}) / N({\rm K~{\small I}}) =  1.1   \pm 0.3 \times 10^{-3}$ 
 toward Sk 143, we obtain a relative abundance of  $A_g({\rm Rb})/A_g({\rm K}) = 0.4 \pm  \times 10^{-3}$. This   is lower than   the solar   value of $2.1 \pm 0.2 \times 10^{-3}$, or equivalently, [Rb/K]=-0.7   \citep{lodders2021SSRv..217...44L}. It is  also lower than that of the interstellar gas     toward $\rho $ Oph A of $(1.3 \pm 0.3) \times 10^{-3}$ and in  the two
 components   toward Cyg OB2   of  $1.4 \times 10^{-3}$ 
and $1.2 \times 10^{-3}$  \citep{Federman2004ApJ...603L.105F}. Because K traces Fe quite well, this result suggests an intrinsic deficiency of Rb in the cloud in front of Sk 143 in general. 

Rubidium has seldom been investigated in stars because it requires access to the near-infrared, and the
spectral range is contaminated by telluric absorption lines that may or may not be blended with the stellar lines, 
depending on the observed radial velocities of the stars.
Rubidium abundances 
were reported  for 19 stars  in the metallicity range $-2.8 <$~[Fe/H]~$< 0$  \citep{Gratton1994A&A...287..927G} and  for 44  giants and dwarfs in the  range of $-2.0<$~[Fe/H]~$< 0.0$  \citep{Tomkin1989MNRAS.241..777T}.  
These  studies showed  that  [Rb/Fe] tends to be moderately supersolar,  $0 \lesssim$~[Rb/Fe]~$\lesssim 0.5$, in  the metal-poor regime ([Fe/H]~$\lesssim -1$) \citep{Lambert1995ApJ...450..302L}.
At variance with these findings, \citet{abia2020A&A...642A.227A}    found   [Rb/Fe] ratios
slightly subsolar by $\approx$  -0.3 dex  at low metallicities. However,  recent studies   confirmed  that  [Rb/Fe] increases with the decrease in metallicity, which is consistent   with the  prediction of chemical evolution models
 \citep{takeda2021AN....342..515T,caffau2021A&A...651A..20C}.
\citet{dorazi2013ApJ...776...59D}  found solar ratios of [Rb/Fe] $\approx$ 0 in a few globular clusters with metallicities [Fe/H] $\approx$ -1.5.

The $^{85}$Rb/$^{87}$Rb isotope ratio toward Sk~143 is found to be   
$^{85}$Rb/$^{87}$Rb = 0.46. This value   differs significantly from 
the meteoritic value of $^{85}$Rb/$^{87}$Rb = 2.430 \citep{lodders2021SSRv..217...44L} and also from  the value  measured in the Galactic interstellar medium  toward $\rho$ Oph A and  HD 169454 of  $^{85}$Rb/$^{87}$Rb =  1.21 and $>$ 2.4, respectively  \citep{kawanomoto_rb_2009ApJ...698..509K}. This suggests  a different importance of r and s-processes, which leads to the Rb isotopes between the MW and the cloud toward Sk~143. The observed isotopic  difference can be obtained either with a lower abundance of $^{85}$Rb or with a higher $^{87}$Rb abundance  relative to the abundances in  meteorites. However,
 the Rb abundance with respect to potassium  can be used  to determine which of the two possibilities is true  \citep{Federman2004ApJ...603L.105F}.  
Toward Sk~143, we find 
$^{85}$Rb/K = $0.6  \times 10^{-3}$ and
$^{87}$Rb/K = $ 2.0  \times 10^{-3}$.  The Solar 
System ratios are $^{85}$Rb/K = $1.38 \pm 0.14 \times 10^{-3}$
and $^{87}$Rb/K = $0.69 \pm 0.06 \times 10^{-3}$ \citep{lodders2021SSRv..217...44L}.  
The comparison   shows that $^{85}$Rb is underabundant in
the gas toward Sk~143 by a factor of about two, and $^{87}$Rb is overabundant by about a factor of four with respect to   the
Solar System abundances.  

Theoretical   models  indicate that $^{87}$Rb mainly arises from the $r$-process 
\citealt{arlandini1999ApJ...525..886A,raiteri1993ApJ...419..207R}).  
We are led to infer that the higher interstellar abundance 
for $^{87}$Rb is due to an  r-process synthesis in the stars that enriched the interstellar gas seen along the line of sight  of Sk~143.

 The relatively low Rb abundance relative to iron suggests a different nucleosynthesis, and    the r-process that occurs in rotating massive metal-poor stars  seems a suitable candidate \citep{limongi2018ApJS..237...13L,choplin2018A&A...618A.133C,banerjee2019ApJ...887..187B}. These  models show that  fast rotation   boosts the r-process. Rotation-induced mixing results in primary production of  $^{13}C$. The presence of $^{13}C$ has been established at low metallicities \citep{molaro2023arXiv230911664M}. $^{13}C$ provides  neutrons via $^{13}C(\alpha ,n) ^{16}O$ during core He burning. In the models of \citet{limongi2018ApJS..237...13L} for a 20 M$\odot$ star with a rotational velocity $\ge$ 150 \kms and [Fe/H] $\le$ -1.0,  $^{87}$Rb  dominates  $^{85}$Rb. Depending on the rotation speed and the mass-loss rate, a strong r-process can occur that produces elements up to bismuth for progenitors with  low metallicities \citep{banerjee2019ApJ...887..187B}. This result suggests that rapidly rotating massive metal-poor stars  can potentially explain the early onset of the r-process we observed in the metal-poor cloud toward Sk~143.

\begin{figure}
        \includegraphics[width=\columnwidth]{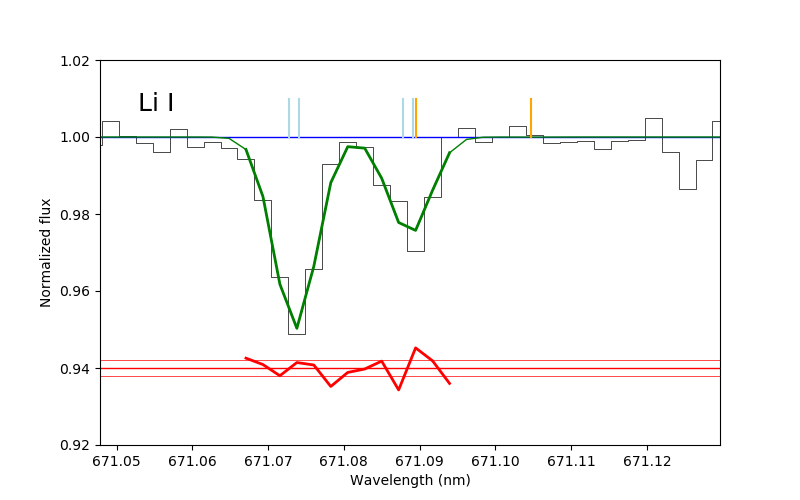}
    \caption{Spectrum of Sk143 around the \lii\  resonance doublet at 670.7 nm. The isotope of \livii\ is plotted in light blue and \livi\ in orange. }
    \label{fig:sk143_li}
\end{figure}

\subsection{Lithium and  \liratio  isotopic ratio in the interstellar medium of the Small Magellanic Cloud}

 \citet{Howk2012Natur.489..121H} detected the interstellar neutral lithium line at 670.7 nm toward Sk~143 and derived a column density of  log N($^7$Li) = $10.29\pm0.02$  and  an 
 isotopic ratio of  $\liratio_{\rm SMC} = 0.13\pm0.05$ with a marginal detection of $^6$Li. 
The ESPRESSO Li~{\small I} region toward Sk~143 is shown in Fig. \ref{fig:sk143_li}.
The estimate of the \liratio\ relies on
simultaneously fitting the hyperfine structure of the \livii\
and \livi\ absorption. The wavelengths for the hyperfine
levels were adopted from \citet{sansonetti1995PhRvA..52.2682S}.
 The stellar continuum  in the region around \LiI\ is featureless and was well fit.  
 In the profile-fitting process,  we allowed variations in the
parameters of the continuum  during the minimization
process.  This allowed us to estimate  the uncertainties caused by
the continuum fitting into the final error budget. 
Because  the absorption is so weak, the \bvalue\ has
a minor effect on the precision  of the column density determination.
We derive  log N(\livii)   = 10.26 $\pm$ 0.02   and
log N($^6$Li) $<$ 9.30 at 3 $\sigma$ C.L..

\subsubsection{$^6$Li}

 \livi\ is not expected to be produced in measurable amounts in
BBN \citep{pospelov2010ARNPS..60..539P}. On the other hand,
 both $^7Li$ 
and \livi\ are directly synthesized in the ISM through cosmic-ray
 interactions with
interstellar medium (ISM) particles, either through spallation or
$\alpha+\alpha$ fusion. These are  the only sources for \livi\ 
post-Big Bang production.
A few years ago, ($\ge 2 \sigma$) \livi\ detections were reported in nine  stars  over a metallicity range of -1.25 $<$ [Fe/H] $<$ -2.74. The stars showed a 
 uniform  \livi/$^7Li$\ ratio as a function of metallicity \citep{asplund2006ApJ...644..229A}. 
These observational determinations imply  a primordial production of 
\livi and no destruction of  $^7Li$  inside of the stars  because stellar interiors destroy
\livi more efficiently   than $^7Li$.
However,   convective processes generate an excess of absorption in the red wing of the \livii\ absorption profile that could be interpreted as  \livi \citep{cayrel2007A&A...473L..37C}.
The reliability of the \livi plateau  was further called into question with a Non-local Thermodynamic Equilibrium   analysis  and new high-quality observations  \citep{lind2013A&A...554A..96L,wang2022MNRAS.509.1521W}. 

Toward Sk~143, we derive   (\liratio)$_{SMC}$ $<$ 0.1. The isotopic ratio is not affected by ionization or dust-depletion effects,  and the limit is robust. The value  is 
consistent with the value for the Solar System,  (\liratio)$_\odot$ = 0.0787 $\pm$0.0004 , and with the (\liratio)$_{MW}$  = 0.13$\pm$0.04 derived for the Milky Way ISM by  \citet{kawanomoto_li_2009ApJ...701.1506K}. 
For standard energy distributions of Galactic cosmic rays, $^7Li$  and \livi\ 
isotopes are produced in a ratio  (\liratio)$_{CR} \approx 0.67$ \citep{steigman1992ApJ...385L..13S}.
Thus, our upper limit  for the  (\liratio)$_{SMC}$ imposes a stringent upper bound of  $<20\%$ to the fraction of the $^7Li$ detected in the  SMC toward Sk~143 that could have been produced via cosmic-ray spallation processes.
 
\subsubsection{$^7$Li}

Determinations of the  $^7$Li abundance in interstellar environments with subsolar metallicities provide a method of constraining the cosmic evolution of $^7$Li  that is not affected by the uncertainties that  hamper stellar $^7$Li abundance studies. The interstellar studies have the potential to provide insights into the well-known discrepancy between the primordial $^7$Li  abundance predicted by standard Big Bang nucleosynthesis and the $^7$Li abundance observed in the atmospheres of metal-poor Galactic halo stars \citep{sbordone2010A&A...522A..26S,mataspinto2021A&A...654A.170M}, the so-called Spite plateau
\citep{spite1982Natur.297..483S,spite1982A&A...115..357S}. The value of  Li produced in the primordial nucleosynthesis mainly depends on the cosmic baryon-to-photon ratio, which can be inferred from the measured primordial deuterium abundance or from observations of the cosmic microwave background. The current primordial \livii\ abundance from BBN plus CMB  is    \abund{^7Li} $\equiv \log [N(^7Li)/N({\rm H})]  +12 = 2.69 \pm 0.14$  \citep{yeh2021JCAP...03..046Y},     \abund{^7Li} = 2.74  $\pm$  0.17 (\citep{pitrrou2021MNRAS.502.2474P} and  \abund{^7Li}  = 2.66 \citep{consiglio2018CoPhC.233..237C}.  These estimates are significantly higher than the  most recent determination of the Li plateau value in the halo stars  of \abund{^7Li}  = 2.16  $\pm$  0.07 by \citet{mataspinto2021A&A...654A.170M}.
This discrepancy could be due to the depletion of stellar lithium abundances or to new physics that were effective in the early Universe.

In the interstellar medium, the derivation of the absolute $^7$Li abundance is complicated by the combined effects of ionization and dust grain depletion. In diffuse clouds,  the  neutral lithium is a trace species, and the dominant ionization state of $^7$Li is $^7$Li~{\sc ii}, which is  not observable. Thus, the observed abundance   requires a large correction for  ionization. However, since   Li~{\sc i} and K~{\sc i} respond in a similar way to   ionization and depletion conditions,    the Li/K ratio can be used  to estimate the amount of photoionization and depletion corrections that are required to infer the absolute $^7$Li abundance \citep{welty2001ApJS..133..345W,knauth2003ApJ...586..268K,white1986ApJ...307..777W}.
The absolute \abund{^7Li}  abundance can be derived through the equation 

\begin{equation}
\label{eq:Li1}
\abund{^7Li}_{\mathrm{SK143}}~=~\abund{^7Li}_{\odot}~+~[K/H]_{\mathrm{SK143}}~+~[^7Li/K]_{\mathrm{SK143}},
\end{equation}

 Along the sight lines through the Milky
Way  $N(\mbox{$^7$LiI\,})/N(\mbox{\KI})$  is found nearly constant, and the ratio is about $-$2.27 \citep{Steigman1996ApJ...457..737S,welty2001ApJS..133..345W}. The constancy of the ratio in the Galaxy  shows that  $^7$Li\,\ and K must  have very
similar ionization and dust depletion behaviors in different environments  of the Milky Way.   Since the solar system is $\log (\mbox{$^7$Li\,}/\mbox{K})_\odot = -1.84\pm 0.05$  Li is more depleted than K with  a  differential ionization correction of $\log~[(\Gamma/\alpha_r)_{\mathrm{Li}}/(\Gamma/\alpha_r)_{\mathrm{K}}]$~=~$+$$0.54\pm0.08$ \citep{Steigman1996ApJ...457..737S,welty2003ApJS..147...61W}. This ionization correction assumes a temperature of $T=100$~K, although the ratio of the recombination coefficients is not very sensitive to $T$ for the species we considered \citep{pequignot1986A&A...161..169P}.
Our measured value of log~$N(^7$Li~{\sc i})/$N$(K~{\sc i})~=~$-$2.31$\pm$ 0.02  implies 
log~$N(^7$Li)/$N$(K)~=~$-$1.77$\pm$0.08 after correction for the differential ionization.   
With the Solar System value  of log~($^7$Li/K)$_{\odot}$~=~$-$1.84$\pm$0.05, we obtain 
$[^7$Li/K]~ $\equiv ~\log $N($^7$Li)/N(K)~$-$~log~($^7$Li/K)$_{\odot}$~=~$+$0.07$\pm$0.08.

The   potassium  abundance   can  be deduced from the zinc abundance    in the same cloud assuming they both track iron closely. \ZnII\ and  \ion{S}{II}  are in their dominant stage, and neither are  depleted, or are depleted only very little, into dust grains. 
However, observations show a mild overabundance of [K/Fe] $\approx$ 0.1 at [Fe/H] = -1 when Non-local Thermodynamic Equilibrium  effects are taken into account for the 769.8 nm line \citep{reggiani2019A&A...627A.177R,takeda2019StGal...2....1T}. 

Thus,  assuming  [Zn/Fe]= 0 and allowing a mild overabundance in [K/Fe] 
 = 0.1  from [Zn/H]$_{143} =  -1.28 \pm 0.06$  , we obtain a     value  of  [K/H]$_{143}\approx -1.18$ 
 for the  potassium abundance in the cloud.  
 This is also consistent with the measured sulphur abundance of  [S/H]=$-0.72 \pm 0.11$, considering that sulphur is an alpha-element and 
  is overabundant by about 0.5 dex with respect to  iron-peak elements at these metallicities
 \citep[see e.g.][and references therein]{Perdigon2021,Lucertini2022}.  In addition, the nature of Zn is not fully understood.
While Zn is often observed to track Fe, at low metallicities, ([Fe/H]$< -0.5$) [Zn/Fe] appears to be constant  and slightly supersolar \citep{duffau2017A&A...604A.128D}. When [Zn/Fe] is supersolar, it decreases the derived iron abundance in the cloud,   and therefore, it also decreases the potassium and lithium abundances. 
 With   a solar Li abundance of $A$(Li)$_{\odot}$~=~$3.27\pm0.03$  \citep{lodders2021SSRv..217...44L}, Eq. \ref{eq:Li1}
becomes

\setlength\arraycolsep{2pt}
\begin{eqnarray}
\abund{^7Li}_{\mathrm{SK143}} &=& 3.27 ~(\pm 0.03) -1.18 ~(\pm0.09) +0.07 ~(\pm0.08)
\label{eq:Li2}
\end{eqnarray}
\begin{eqnarray}
\abund{^7Li}_{\mathrm{SK143}}  &=& 2.16~ (\pm0.12)
\label{eq:Li3}
\end{eqnarray}

 The  value is closer to the Li value observed in the halo stars rather than the BBN+CMB  value. 
\citet{decia2024A&A...683A.216D}  estimated  the metallicity of the  material toward Sk 143   by considering several elements and including the effect  of the presence of dust. They derived a  total metallicity  of  [M/H] = -1.09 $\pm$ 0.24, which   assumes that  [K/H]=[M/H] provides an absolute lithium abundance of A(Li)= 2.25$\pm$ 0.25. 
\citet{Howk2012Natur.489..121H} took the K abundance from  stellar determinations, which is higher by about 0.7 dex, taken from the stars, and therefore, computed a much higher value of \abund{^7Li}. The metallicity distribution of stars belonging to the field FLD-121, like the gas toward Sk~143,  shows a clear lack of relatively metal-rich stars with [Fe/H] between –0.8 and –0.5 dex and peaks 
at $\approx$ -1.1 \citep{mucciarelli2023A&A...671A.124M}. It is also important to note that at these metallicities, no significant stellar
production of Li is expected in the SMC in the detailed models by \citep{izzo2022MNRAS.510.5302I}, which are based on $^7$Be detection in two SMC novae. A measured value about \abund{^7Li} $\approx$ 2.2 is a quite universal value in old stars of the Galactic halo, in the metal-poor stars belonging to the accreted galaxy  of Gaia-Sausage-Enceladus \citep{molaro2020MNRAS.496.2902M}, and in the stars of the Sagittarius galaxy \citep{mucciarelli2014MNRAS.444.1812M} and $\omega$ Cen \citep{monaco2010A&A...519L...3M}, which is  very likely the core of a disrupted galaxy.
This strongly suggests that the Spite plateau is universal, as is the cosmological lithium problem, namely  the discrepancy between
observed Li and the predictions of standard Big Bang nucleosynthesis.  The generally favoured solution of this problem involves  Li depletion by several mechanisms in the stellar atmospheres
\citep[see][for a recent discussion]{fields2022JCAP...10..078F}. However, while the theory of turbulent diffusion \citep{richard2002ApJ...580.1100R,richard2005ApJ...619..538R,borisov2024arXiv240315534B} may be invoked to explain this discrepancy in stars, it cannot have any effect on the $^7$Li as measured in low-metallicity  interstellar gas, and  our  measurement in metal-poor gas toward Sk~143 in the SMC challenges the stellar depletion  solution.  Some alternative solutions with new physics  are described in \citet{mathews2020MmSAI..91...29M}.

\section{Conclusions}\label{sec13}

Based on spectrum with a high resolution and high signal-to-noise ratio obtained with the ESPRESSO spectrograph, we searched for the presence of rare elements  in  the  gas along the line of sight of  the star Sk~143 in the SMC. The absorbing gas was observed at a heliocentric velocity of 131 \kms and   originated in a  low-metallicity substructure   of the SMC
\citep{molaro1990MmSAI..61..677M,muciarelli2023A&A...677A..61M}. This search led to the main findings we list below.

\begin{itemize}

\item{The resonance lines of Rb were detected  outside the Galaxy for the first time. Furthermore,  the lines of the  two  isotopes  $^{85}$Rb and $^{87}$Rb were resolved. The isotopic ratio is  the opposite  to the ratio that is observed  in the  Galaxy. By using  potassium as a reference, we showed that the observed ratio  is due  to an enhancement   of $^{87}$Rb   and  to a deficit of $^{85}$Rb. This suggests that the 
{\it r}-process dominates   in the  synthesis of rubidium.  Using  Zn as a proxy for Fe,  we  derived   [Rb/Fe] =-0.55. This is  lower than that  observed in the Galaxy and than what is expected in the SMC stars. 
These two anomalies  favor the possibility that $r-$ elements are produced in metal-poor rotating massive stars \citep{limongi2018ApJS..237...13L,banerjee2019ApJ...887..187B}.
}

\item{  No evidence was found of the presence of the $^6$Li isotope. The upper bound is slightly below the marginal detection  of  \citet{Howk2012Natur.489..121H}}.

\item{The presence and abundance of $^7$Li   was found to agree with \citet{Howk2012Natur.489..121H}. However,   we measured    the metallicity of the cloud  with zinc and found  [Zn/H]=-1.3, which agrees with the metallicities   of the stars    of  the  SMC component that show the same radial velocity \citep{mucciarelli2023A&A...671A.124M}. This leads to an absolute  lithium of   \abund{^7Li}  $\approx$  2.2,  which  is about   the value measured in warm-halo  dwarf stars.  This  suggests that the  solution of the cosmological Li problem   cannot be ascribed to stellar depletion.}
\end{itemize}

The line of sight toward the hot star Sk~143 in the Small Magellanic Cloud proves to be unique because it offers  the possibility of studying the abundance of elements that are inaccessible to extragalactic investigations.  In particular, it offers insight into the  the values of isotopic ratios in  a low-metallicity environment, which provides  unique information on the processes of their chemical synthesis and  evolution.

\begin{acknowledgements}
  
The entire ESPRESSO team is gratefully thanked for having built a high quality and precise instrument. We also thank the ESO staff for having assisted and conducted the observations in service mode with great professionalism. Stimulating discussions with 
Marco Limongi, Serrgio Cristallo, Gabriele Cescutti, Ed Jenkins  and Ryan Cooke are acknowledged. Gabriella Schiulaz is warmly thanked for  checking the English.  Funding by the European Union – NextGenerationEU RFF M4C2 1.1 PRIN 2022 project "2022RJLWHN URKA is also acknowledged.
   
\end{acknowledgements}

\bibliography{sk143}

\begin{appendix}
\section{Column densities toward Sk~143}
\begin{sidewaystable*}
\caption{ Column densities toward Sk~143. }
\label{tab:full}
\begin{tabular*}{\textheight}{@{\extracolsep\fill}lrlcccrcc}
\hline
\hline
Species &$\lambda$ & $f$-value  & $b$ & Vr &$\log N$ & $\log N_{Howk}$ &  $\abund{X}_\odot$ & [X/H]$_{\rm SMC}$\\
 & (\AA) & & \kms  & \kms  & (cm$^{-2}$)&   (cm$^{-2}$) &&\\ 
\hline
\HI           &  &&&&& 21.07 $\pm$ 0.05 &  $\equiv12.00$ &  \\
H$_{2}$         &  &&&&& 20.93 $\pm $ 0.09 &  $\equiv 12.00$ & \\
H$_{\rm total}$ &&& & & &21.46 $\pm$ 0.04 & $\equiv12.00$ &  \\
\ion{S}{II}   &1250.578 & 0.00543&1.6&133.62 $\pm$ 0.09&$15.98\pm0.11$&$\ga15.17\pm0.05$ &  $7.24\pm0.03$ & $-0.72 \pm 0.11$ \\
\ion{S}{II}   &1253.805 & 0.0109 &1.6&133.62 $\pm$ 0.09 &$15.98\pm0.11$&$\ga15.17\pm0.05$ &  $7.24\pm0.03$ & $-0.72 \pm 0.11$ \\
 \ion{S}{II}   & 1250.584& & 1.6& 140.69 $\pm$ 0.09&$14.68\pm0.05$& $-$ &   &  \\
  \ion{S}{II}   & 1253.811& &1.6&140.69 $\pm$ 0.09&$14.68\pm0.05$& $-$ &   &  \\
\ZnII         & 2026.1370 &0.630$^c$ &1.6 &131.42 $\pm$ 0.15&$12.77 \pm0.09$&$\ga12.80\pm0.06$ &  $4.56\pm0.03$ & $-1.28\pm0.09$ \\
\ZnII         & 2062.6604 & 0.309$^c$ &1.6 &131.42 $\pm$ 0.15&$12.77 \pm0.09$&$\ga12.80\pm0.06$ &  $4.56\pm0.03$ & $-1.28\pm0.09$ \\
\ion{Mg}{I}  & 2026.48 & 0.113 & 1.0 & 131.67 $\pm$ 0.06& 14.28  $\pm 0.29$ &  &7.56 $\pm 0.05$ & $-2.74 \pm 0.07$\\
\KI           & 4044.143  &0.00609  $^b$ &1.0& $132.60 \pm 0.06$ &$12.57 \pm 0.10$ & $12.61 \pm 0.02$ &  $5.04\pm 0.04$ & $-1.93\pm0.04$ \\
\NaI          & 5889.951& 0.641 &1.6&131.91$ \pm$ 0.03&14.33 $\pm$ 0.02  & $>12.96\pm0.04$   &  $6.36\pm0.03$ & $-1.68$$ \pm $ 0.05 \\ 
\NaI          & 5895.924&0.320 &1.6&131.91$ \pm$ 0.03&14.33 $\pm$ 0.02  & $>12.96\pm0.04$   &  $6.36\pm0.03$ & $-1.68$$\pm $ 0.05 \\ 
\NaI          &5889.951&0.641& 1.6 & 138.91$ \pm$ 0.03              & 10.81 $\pm$  0.05  &  $-$                &                 & \\
\NaI          &5895.924&0.320& 1.6 & 138.91$ \pm$ 0.03              & 10.81 $\pm$  0.05 &  $-$                &                 & \\
\ion{$^7$Li}{I}      & 6707.754& 0.186 &1.0 & &  &  &  &  \\
\ion{$^7$Li}{I}      &6707.766& 0.309 & 1.0 & &  &  &  &  \\
\ion{$^7$Li}{I}      &6707.904& 0.093 &1.0  & &  &  &  &  \\
\ion{$^7$Li}{I}      &6707.917& 0.155& 1.0 & &  &  &  &  \\
\ion{$^7$Li}{I}      &&1.0& & 132.93 $\pm$ 0.05&10.26$\pm$ 0.02 & 10.29 $\pm$ 0.02 &  $3.27\pm0.03$ & $-2.7\pm0.04$ \\
\ion{$^6$Li}{I}      &6707.921& 0.495& 1.0 & &  &  &  &  \\
\ion{$^6$Li}{I}      &6708.072& 0.247& 1.0 & &  &  &  &  \\
\ion{$^6$Li}{I} &&1.0&1.6&&$<$9.30 & $9.41\pm 0.15 $  &  2.14 $\pm$ 0.05 &  \\
\ion{$^{85}$Rb}{I}& 7800.232  &$0.290 $\ $^a$ &1.0&&9.46 $\pm$ 0.13 && & \\
\ion{$^{85}$Rb}{I}& 7800.294  &$0.406 $\ $^a$ &1.0&&9.46 $\pm$ 0.13 && & \\
\ion{$^{87}$Rb}{I}&  7800.183 & $0.261 $\ $^a$&1.0&& 9.80 $\pm$ 0.06 && & \\
\ion{$^{87}$Rb}{I}&  7800.321 & $0.435 $\ $^a$&1.0&& 9.80 $\pm$ 0.06 && & \\
\ion{Rb}{I} &  &1.0&&132.27 $\pm$ 0.18 & 9.60 $\pm$ 0.08 & $-$ &2.36 $\pm$ 0.03 & $-1.86$ $\pm$ 0.09 \\
\hline
\end{tabular*}
\tablefoot{The solar abundances are taken from  \citet{lodders2021SSRv..217...44L}. (a) \citet{morton2000ApJS..130..403M}  (b) \citet{morton1991ApJS...77..119M}, c) \citet{kisielius2015ApJ...804...76K}. The wavelengths for the lines in the STIS spectrum are in vacuum, and those in the ESPRESSO spectrum are in air.}
\end{sidewaystable*}
\end{appendix}
\end{document}